\documentclass[a4paper]{article}

\begin{document}

\title {\bf Quantum Stationary Hamilton Jacobi \\ Equation
in 3-D for Symmetrical Potentials. \\Introduction of the
Spin.}

\author{T.~Djama\thanks{E-mail:
{\tt djam\_touf@yahoo.fr}}}

\maketitle
\begin{center}Universit\'e des Sciences et de la
Technologie
Houari Boum\'edienne, \\Alger, Algeria \\ 
\vspace*{0.5cm}
Mail address:14, rue Si El Hou\`es,
B{\'e}ja{\"\i}a, Algeria
\end{center}

\begin{abstract}
\noindent
We establish the quantum stationary Hamilton-Jacobi
equation in
3-D and its solutions for three symmetrical
potentials: Cartesian
symmetry potential, spherical symmetry potential and
cylindrical
symmetry potential. For the two last potentials, a new
interpretation
of the spin is proposed within the framework of
trajectory representation.
\end{abstract}

\vskip\baselineskip



\newpage
 %
\vskip\baselineskip
\noindent
{\bf 1- Introduction}
\vskip\baselineskip
%

Since twenty years, the trajectory representation has
been
proposed by Floyd as a new approach of quantum
mechanics.
In fact, Floyd took up the QSHJE in one dimension,
solved it
\cite{Floyd1,Floyd2,Floyd3}
and constructed quantum trajectories from the Jacobi's
theorem
\cite{Floyd4,Floyd5}
\begin{equation}
{\partial S_0 \over \partial E}=t-t_0\; .
\end{equation}
After fifteen years, Faraggi and Matone derived one
dimensional QSHJE from the equivalence postulate
\cite{FM1,FM2,FM3,FM4}.
They took up the Jacobi's theorem and established an
equation
of quantum motion \cite{FM3}. Then with Bertoldi
\cite{FM3,FM4},
they derived the N-dimensional QSHJE. Later, Bouda
\cite{Bouda}
re-investigated the QSHJE and proposed to write the
reduced action as
\begin{equation}
S_0=\hbar \arctan\left({\mu\theta+\phi \over
\theta+\nu \phi}\right)
+e\ \hbar\; ,
\end{equation}
where $\mu$, $\nu$ and $e$ are real constants
identified
to the integration constants of the QSHJE.

Recently, Bouda and Djama have criticized the use of
Jacobi's 
theorem as written in Eq. (1), to
derive trajectories equation since this theorem is
applied in
classical mechanics for a first order differential
equation
while the QSHJE is a third order one \cite{BD1}. They
have
constructed, in one dimension, a quantum Lagrangian
and
formulated a quantum version of Jacobi's theorem from
which
they obtained the relation connecting velocity
$\dot{x}$
and conjugate momentum $\partial S_0 / \partial x$.
Then, in
a recent paper \cite{BD2}, Bouda and Djama have
plotted quantum
trajectories for several potentials, and have given a new
physical
meaning of the de Broglie wavelength.
 
The construction of the Hamilton-Jacobi equation in
one dimension
within the framework of the quantum mechanics
\cite{Floyd1,Floyd4},
is an important step for the elaboration of the
deterministic quantum
theory introduced by Floyd
\cite{Floyd1,Floyd2,Floyd3,Floyd4},
and reformulated by Bouda and
Djama \cite{BD1,BD2}. Nevertheless, such equation is
not
sufficient for an adequate description of reality.
Indeed, physical
phenomena are often realized in three dimensional
space.
For this reason, a generalization of the quantum
stationary
Hamilton-Jacobi equation (QSHJE) into 3-D prove to be
indispensable. For this purpose, we introduce an
attempt of such
generalization. Thus, we construct in this letter the
3-D
QSHJE for three symmetrical potentials. In Sec. 2, the
Cartesian
symmetry potential is studied. Then, in Sec. 3 we
investigate
the case of the spherical symmetry potential. After,
in Sec. 4,
we approach the cylindrical symmetry potential.
Finally, in Sec. 5, we comment on our results and
propose a
new interpretation of the spin and its physical
meaning in trajectory
interpretation of quantum mechanics.

\newpage

%
\vskip\baselineskip
\noindent
{\bf 2- Cartesian Symmetry Potential}
\vskip\baselineskip
%

Let us begin by a potential with Cartesian symmetry
whose
expression is given by
\begin{equation}
V(\vec{r})=V_{x}(x)+V_{y}(y)+V_{z}(z)=\sum_{q}V(q)\ \
\;;q=x,y,z\;.
\end{equation}
An example of such a potential is the spatial
harmonic
oscillator's potential. The stationary Schr\"{o}dinger
equation is
\begin{equation}
-{\frac{\hbar ^{2}}{2m}}\Delta \psi +V(\vec{r})\;\psi
(\vec{r})=E\ \psi (%
\vec{r})\;.
\end{equation}
For a potential given by Eq. (3), the wave function is
written as the
product of three functions of one variable
\begin{equation}
\psi (\vec{r})=\prod_{q}\phi_{q}(q)\; ;\ \ \ \ \
q=x,y,z\;,
\end{equation}
\noindent
and the corresponding energy is
\begin{equation}
E=\sum_{q}E_{q}\; ;\ \ \ \ \ q=x,y,z\;.
\end{equation}
By substituting Eqs. (3), (5) and (6) in Eq. (4), one
get to the one
dimensional Schr\"{o}dinger equations
\begin{equation}
-{\frac{\hbar ^{2}}{2m}}{\frac{d^{2}\phi
_{q}}{dq^{2}}}+V_{q}(q)\phi
_{q}(q)=E_{q}\phi _{q}(q)\;,\ \ \ \ \ q=x,y,z\;.
\end{equation}
As in the one dimensional approach \cite{FM1,FM2}, let
us write
the functions $\phi _{q}$ in the following form
\begin{equation}
\phi _{q}(q)=A_{q}(q)\left[ \alpha _{q}\
e^{{\frac{i}{\hbar }}\
S_{0q}(q)}+\beta _{q}\ e^{-{\frac{i}{\hbar }}\
S_{0q}(q)}\right] \;,\ \ \ \
\ q=x,y,z\;,
\end{equation}
where $A_{q}(q)$ and $S_{0q}(q)$ are real functions.
$\alpha _{q}$ and $\beta _{q}$ are complex constants.
Note that the choice of expressions (8) for the
function
$\phi _{q}$ is justified in Refs. \cite{FM1,FM2,Bouda}
where
the same form as (8) is used for the one dimensional
wave
function $\psi (x)$. Then, it is obvious to take up
this form
for each one dimensional function $\phi _{q}$.
Taking Eq. (8) into Eq. (7), we obtain
\begin{eqnarray}
\left[ {\frac{1}{2m}}\left( {\frac{\partial
S_{0q}(q)}{\partial q}}\right)
^{2}-{\frac{\hbar ^{2}}{2m}}{\frac{\partial
^{2}A_{q}/\partial q^{2}}{A_{q}}}%
+V_{q}-E_{q}\right] \left( \alpha _{q}\
e^{{\frac{i}{\hbar }}\
S_{0q}(q)}\right. \ \ \ \ \ \ \ \ \ \ \ \ \ \ \ \ \ \
\hskip5mm &&
\nonumber \\
\left.+\beta _{q}\ e^{-{\frac{i}{\hbar }}\
S_{0q}(q)}\right)
-{\frac{i\hbar }{2m\ A_{q}^{2}}}{\frac{\partial
}{\partial q}}\left(
A_{q}^{2}{\frac{\partial S_{0q}}{\partial q}}\right)
\left( \alpha _{q}\ e^{{%
\frac{i}{\hbar }}\ S_{0q}(q)}-\beta _{q}\
e^{-{\frac{i}{\hbar }}\
S_{0q}(q)}\right)=0 \; ,
\end{eqnarray}
which is satisfied when the two following equations
are so
\begin{equation}
{\frac{1}{2m}}\left( {\frac{\partial
S_{0q}(q)}{\partial q}}\right) ^{2}-{%
\frac{\hbar ^{2}}{2m}}{\frac{\partial
^{2}A_{q}/\partial q^{2}}{A_{q}}}%
+V_{q}-E_{q}=0\;,
\end{equation}
\begin{equation}
{\frac{\partial }{\partial q}}\left(
A_{q}^{2}{\frac{\partial S_{0q}}{%
\partial q}}\right) =0\;.
\end{equation}

Now, as it is the case in one dimension, Eqs (10) and
(11) lead to
\begin{equation}
{\frac{1}{2m}}\left( {\frac{\partial
S_{0q}(q)}{\partial q}}\right) ^{2}-{%
\frac{\hbar ^{2}}{2m}}\left\{ S_{0q},q\right\}
+V_{q}=E_{q}\;,
\end{equation}
where $\{ S_{0q},q\} $ is the Schwarzian derivative of
$S_{0q}$
with respect to $q$ and given by
\begin{equation}
\left\{ S_{0q},q\right\} =\left[ {\frac{3}{2}}\left(
{\frac{\partial
^{2}S_{0q}/\partial q^{2}}{\partial S_{0q}/\partial
q}}\right) -{\frac{%
\partial ^{3}S_{0q}/\partial q^{3}}{\partial
S_{0q}/\partial q}}\right] \;,
\end{equation}
and $\partial S_0 / \partial q$ representing the
conjugate
 momentum along the direction $q$. 

\noindent
Eqs. (12) represents the components of the QSHJE along
each
of the three directions $x$, $y$ and $z$. If we set
\begin{equation}
S_{0}=\sum_{q}S_{0q}\;,\ \ \ \ \ q=x,y,z,
\end{equation}
where $S_{0}$ define the reduced action in 3-D, then
summing the 
three Eqs. (12) and taking in account Eqs. (3), (6)
and (14),
we find
\begin{equation}
{\frac{1}{2m}}(\vec{\nabla}S_{0})^{2}-{\frac{\hbar
^{2}}{4m}}\sum_{q}\left\{
S_{0q},q\right\} +V(\vec{r})=E\;.
\end{equation}
Eq. (15) represents the QSHJE in 3-D for a potential
with
Cartesian symmetry. It is a third order differential
equation
since the Schwarzian contains the third derivative of
$S_{0}$
with respect to $q$. The terms of more than one order
of the
derivative with respect to $q$ are proportional to
$\hbar ^{2}$
meaning that at the classical limit $(\hbar \to 0)$
these terms vanish and Eq. (15) goes to the classical
stationary Hamilton-Jacobi equation (CSHJE)
\begin{equation}
{\frac{1}{2m}}(\vec{\nabla}S_{0})^{2}+V(\vec{r})=E\;.
\end{equation}
These results are already presented by Faraggi
and Matone \cite{FM1,FM3} within the framework of the
equivalence postulate and using differential geometry.

\noindent
It is useful to remark that through our equations
for Cartesian symmetry potential, a dynamical approach
of the motion, as it is done in one dimension
\cite{BD1}, seems to be possible by treating each
one of the variables $q$ separately from the two
others

\vskip\baselineskip
\noindent
{\bf 3. The Spherical Symmetry Potential}
\vskip\baselineskip
%

Let us now examine the case of the spherical symmetry
potential. For this purpose, we take up the stationary
Schr\"{o}dinger equation written with spherical
coordinates
\begin{eqnarray}
{\frac{\partial ^{2}\psi }{\partial
r^{2}}}+{\frac{2}{r}}{\frac{\partial
\psi }{\partial r}}+{\frac{1}{r^{2}}}{\frac{\partial
^{2}\psi }{\partial
\vartheta ^{2}}}+\hskip5mm &&  \nonumber \\
{\frac{\cot \vartheta }{r^{2}}}\ {\frac{\partial \psi
}{\partial \vartheta }}%
+{\frac{1}{r^{2}\sin ^{2}\vartheta }}\ {\frac{\partial
^{2}\psi }{\partial
\varphi ^{2}}}+{\frac{2m}{\hbar ^{2}}}[E-V(r)]\psi
&=&0\;,
\end{eqnarray}
for which the potential depends only on the radius $r$,
$V(\vec{r})=V(r)$. The Schr\"{o}dinger wave function
can be written in the following form \cite{Tannoudji}
\begin{equation}
\psi (\vec{r})=R(r)\ T(\vartheta )\ F(\varphi )\;,
\end{equation}
where $R(r)$ is the radial wave function, $T(\vartheta
)$
and $F(\varphi )$ are the angular wave functions.  An
example
of such a problem is the dynamics of an electron in
hydrogeno\"{\i}d
atoms. If one substitute Eq. (18) in Eq. (17), one
gets to the following relations \cite{Tannoudji}
\begin{equation}
{\frac{r^{2}}{R}}{\frac{d^{2}R}{dr^{2}}}+{\frac{2r}{R}}{\frac{dR}{dr}}+{%
\frac{2mr^{2}}{\hbar ^{2}}}(E-V(r))=\lambda \;,
\end{equation}
\begin{equation}
{\frac{d^{2}T}{d\vartheta ^{2}}}+\cot \vartheta
{\frac{dT}{d\vartheta }}%
+\left( \lambda -{\frac{m_{\ell }^{2}}{\sin
^{2}\vartheta }}\right) T=0\;,
\end{equation}
\begin{equation}
{\frac{d^{2}F}{d\varphi ^{2}}}+m_{\ell }^{2}\ F=0,
\end{equation}
where $\lambda $ is a constant such as
$$
\lambda =\ell (\ell +1),
$$
$\ell $ being positive integer or vanish, and $m_{\ell
}$ is an integer
satisfying to
$$
-\ell \leq m_{\ell }\leq {\ell }\;.
$$
$\ell (\ell +1)$ is the eighen value of $L^{2}$, $L$
representing the angular momentum operator.
$m_{\ell }$ represents the eighen value of $L_{z}$,
the projection operator of $L$ along the $z$ axis.

Let us write Eqs. (19), (20) and (21) in a form
analogous
to the one dimensional Schr\"{o}dinger equation (Eq.
(7)).
First, take up Eq. (19) and write that
\begin{equation}
R(r)={\frac{\mathcal{X}(r)}{r}}\;,
\end{equation}
where $\mathcal{X}(r)$ is a function of the radius.
Then by taking twice the derivative of Eq. (22) with
respect
to $r$, we obtain
\begin{equation}
{\frac{dR}{dr}}={\frac{1}{r}}{\frac{d\mathcal{X}}{dr}}-{\frac{\mathcal{X}}{%
r^{2}}}\ ,
\end{equation}
\begin{equation}
{\frac{d^{2}R}{dr^{2}}}={\frac{1}{r}}{\frac{d^{2}\mathcal{X}}{dr^{2}}}-2{%
\frac{d\mathcal{X}/dr}{r^{2}}}+2{\frac{\mathcal{X}}{r^{3}}}\
.
\end{equation}
After substituting Eqs. (22) and (23) in Eq. (19), one
gets
\begin{equation}
{\frac{-\hbar
^{2}}{2m}}{\frac{d^{2}\mathcal{X}}{dr^{2}}}+\left[
V(r)+{\frac{%
\lambda \hbar ^{2}}{2mr^{2}}}\right]
\mathcal{X}=E\mathcal{X}\ .
\end{equation}
This last equation has the same form as the one
dimensional Schr\"{o}dinger equation for
$\mathcal{X}(r)$
with an energy $E$ and a fictive potential
\begin{equation}
V^{^{\prime }}(r)=V(r)+{\frac{\lambda \hbar
^{2}}{2mr^{2}}}\ .
\end{equation}
Likewise, the form of Eq. (20) can be reduced to the
form of Eq. (7). This can be done by introducing the
function $\mathcal{T}(\vartheta )$ defined as
\begin{equation}
\mathcal{T}(\vartheta )=\sin ^{\frac{1}{2}}\vartheta
T(\vartheta )\ .
\end{equation}
Taking twice the derivative of this last equation with
respect to $\vartheta$, we deduce
\begin{equation}
{\frac{dT}{d\vartheta }}={\frac{1}{\sin
^{\frac{1}{2}}\vartheta }}{\frac{d%
\mathcal{T}}{d\vartheta }}-{\frac{1}{2}}{\frac{\cos
\vartheta }{\sin ^{\frac{%
3}{2}}\vartheta }}\mathcal{T}\ ,
\end{equation}
\begin{equation}
{\frac{d^{2}T}{d\vartheta ^{2}}}={\frac{1}{\sin
^{\frac{1}{2}}\vartheta }}{%
\frac{d^{2}\mathcal{T}}{d\vartheta ^{2}}}-{\frac{\cos
\vartheta }{\sin ^{%
\frac{3}{2}}\vartheta
}}{\frac{d\mathcal{T}}{d\vartheta }}+{\frac{3}{4}}{%
\frac{\cos ^{2}\vartheta }{\sin
^{\frac{3}{2}}\vartheta }}+{\frac{1}{2}}{%
\frac{\mathcal{T}}{\sin ^{\frac{1}{2}}\vartheta }}\ .
\end{equation}
If we replace Eqs. (28) and (29) in Eq. (20), we obtain
\begin{equation}
{\frac{d^{2}\mathcal{T}}{d\vartheta ^{2}}}+(\lambda
+{\frac{1}{4}})\mathcal{T%
}+{\frac{(1/4-m_{\ell }^{2})}{\sin ^{2}\vartheta
}}\mathcal{T}=0\ .
\end{equation}
Remark that Eq. (30) has the form of the one
dimensional
Schr\"{o}dinger equation with a potential
\begin{equation}
V(\vartheta )={\frac{\hbar ^{2}}{2m}}{\frac{(m_{\ell
}^{2}-1/4)}{\sin
^{2}\vartheta }}\ ,
\end{equation}
and an energy
\begin{equation}
E_{\vartheta }=(\lambda +{\frac{1}{4}}){\frac{\hbar
^{2}}{2m}}\ .
\end{equation}
Remark also that Eq. (21) has the same form as Eq. (7)
with
a vanishing potential and an energy equal to
$(m_{\ell }^{2}\hbar ^{2}/2m)$.

\noindent
Because Eqs. (19), (20) and (21) come to the form of
the one
dimensional Schr\"{o}dinger equation, it is legitimate
to take up
the form (8) for the function $\mathcal{X}(r)$,
$\mathcal{T}(\vartheta )$ and $F(\varphi )$ so to
write them as
\begin{equation}
\mathcal{X}(r)=A(r)\left[ \alpha e^{{\frac{i}{\hbar
}}Z(r)}+\beta e^{-{\frac{%
i}{\hbar }}Z(r)}\right] \ ,
\end{equation}
\begin{equation}
\mathcal{T}(\vartheta )=\xi (\vartheta )\left[ \gamma
e^{\frac{i}{\hbar }%
L(\vartheta )}+\varepsilon e^{-{\frac{i}{\hbar
}}L(\vartheta )}\right] \ ,
\end{equation}
\begin{equation}
F(\varphi )=\eta (\varphi )\left[ \sigma
e^{{\frac{i}{\hbar }}M(\varphi
)}+\omega e^{-{\frac{i}{\hbar }}M(\varphi )}\right] \
.
\end{equation}
By replacing Eqs. (33), (34) and (35) respectively in
Eqs. (25), (30) and (21), we get
\begin{equation}
{\frac{1}{2m}}\left( {\frac{dZ}{dr}}\right)
^{2}-{\frac{\hbar ^{2}}{2m}}{%
\frac{1}{A}}{\frac{d^{2}A}{dr^{2}}}+V(r)+{\frac{\lambda
\hbar ^{2}}{%
2m\;r^{2}}}=E\ ,
\end{equation}
\begin{equation}
{\frac{d}{dr}}\left( A^{2}\ {\frac{dZ}{dr}}\right)=0
\; ,
\end{equation}
\begin{equation}
\left( {\frac{dL}{d\vartheta }}\right)
^{2}-{\frac{\hbar^2}{\xi }}{\frac{d^{2}\xi
}{d\vartheta ^{2}}}-(\lambda +{\frac{1}{4}})\hbar
^{2}-{\frac{(1/4-m_{\ell
}^{2})}{\sin ^{2}\vartheta }}\hbar ^{2}=0\ ,
\end{equation}
\begin{equation}
{\frac{d}{d\vartheta }}\left( \xi ^{2}\
{\frac{dL}{d\vartheta }}\right)=0 \;,
\end{equation}
\begin{equation}
\left( {\frac{dM}{d\varphi }}\right) ^{2}-{\frac{\hbar
^{2}}{\eta (\varphi )}%
}{\frac{d^{2}\eta }{d\varphi ^{2}}}-m_{\ell }^{2}\hbar
^{2}=0\;,
\end{equation}
\begin{equation}
{\frac{d}{d\varphi }}\left( \eta ^{2}\
{\frac{dM}{d\varphi }}\right)=0 \;.
\end{equation}
As in one dimensional case, these equations lead to
\begin{equation}
{\frac{1}{2m}}\left( {\frac{dZ}{dr}}\right)
^{2}-{\frac{\hbar ^{2}}{4m}}%
\left\{ Z,r\right\} +V(r)+{\frac{\lambda \hbar
^{2}}{2mr^{2}}}=E\ ,
\end{equation}
\begin{equation}
\left( {\frac{dL}{d\vartheta }}\right)
^{2}-{\frac{\hbar ^{2}}{2}}\left\{
L,\vartheta \right\} +{\frac{(m_{\ell }^{2}-1/4)}{\sin
^{2}\vartheta }}\hbar
^{2}=(\lambda +{\frac{1}{4}})\hbar ^{2}\;,
\end{equation}
\begin{equation}
\left( {\frac{dM}{d\varphi }}\right) ^{2}-{\frac{\hbar
^{2}}{2}}\left\{
M,\varphi \right\} =m_{\ell }^{2}\hbar ^{2}\ .
\end{equation}
Eqs. (42), (43) and (44) represent the components of
the
QSHJE in 3-D for a spherical symmetry potential.
Note that these equations contain the Schwarzian
derivatives
of the functions $\mathcal{X}$, $\mathcal{T}$ and $F$
respectively, and the conjugates momentums $dZ/dr$,
$dL/d\vartheta$ and $dM/d\varphi$. Now, by deducing
the
value of $\lambda $ from (42) and the value of
$m_{\ell }$
from (44), then substituting in Eq. (43), one obtain
\begin{eqnarray}
{\frac{1}{2m}}\left[ \left( {\frac{dZ}{dr}}\right)
^{2}+{\frac{1}{r^{2}}}%
\left( {\frac{dL}{d\vartheta }}\right)
^{2}+{\frac{1}{r^{2}\sin
^{2}\vartheta }}\left( {\frac{dM}{d\varphi }}\right)
^{2}\right] -\hskip5mm
&&  \nonumber \\
{\frac{\hbar ^{2}}{4m}}\left[ \left\{ Z,r\right\}
+{\frac{1}{r^{2}}}\left\{
L,\vartheta \right\} +{\frac{1}{r^{2}\sin
^{2}\vartheta }}\left\{ M,\varphi
\right\} \right] +\hskip5mm &&  \nonumber \\
V(r)-{\frac{\hbar ^{2}}{8m\ r^{2}}}-{\frac{\hbar
^{2}}{8m\ r^{2}\sin
^{2}\vartheta }} &=&E\ .
\end{eqnarray}
Let us define the reduced action in this case as
$$
S_{0} (r,\vartheta ,\varphi )=Z(r)+L(\vartheta
)+M(\vartheta )\ .
$$
Taking up this equation in Eq. (45), we find
\begin{eqnarray}
{\frac{1}{2m}}  \left( \vec{\nabla}_{r,\vartheta
,\varphi }S_{0}\right) ^{2}-
{\hbar ^{2} \over 4m} \left[ \left\{ S_{0},r\right\}
+{\frac{1}{r^{2}}}\left\{ S_{0},\vartheta
\right\}
\right. \hskip5mm &&
\nonumber \\
\left.  +{\frac{1}{%
r^{2}\sin ^{2}\vartheta }}\left\{ S_{0},\varphi
\right\} \right] +V(r)-{%
\frac{\hbar ^{2}}{8m\ r^{2}}}-{\frac{\hbar ^{2}}{8m\
r^{2}\sin ^{2}\vartheta
}} &=&E\ .
\end{eqnarray}
Eq. (46) represents the QSHJE in 3-D for a spherical
symmetry
potential. At the classical limit $(\hbar \to 0)$, Eq.
(46) goes to
\begin{equation}
{\frac{1}{2m}}\left( \vec{\nabla}_{r,\vartheta
,\varphi }S_{0}\right)
^{2}+V(r)=E\ .
\end{equation}
which is the CSHJE. Note that taking the classical
limit
in Eqs. (42), (43) and (44) making $dL/d\vartheta $ and
$dM/d\varphi $ vanish, then, one cannot obtain the
CSHJE (Eq.(47)). The classical limit must be taken in
Eq. (46).
Note also that considering the QSHJE, and separate
variables
we get to Eqs. (42), (43)
and (44) which lead to Eqs. (19), (20) and (21). After
separating
variables in Eq. (46), three constants of motion
appear.
They are the energy $E$, $\lambda $ and $m_{\ell }$.
These
two last constants will be linked in Sec. 5 with the
quantum
angular momentum.

Before introducing the solutions of Eqs. (42), (43)
and (44),
remark that these equations are identical to the one
dimensional
QSHJE with particular potentials and energies. These
three
equations are solvable when Eqs. (19), (20) and (21)
are so.
Then, as the solution of the one dimensional QSHJE is
written
in terms of two independent solutions of the
stationary
Schr\"odinger equation, the solutions of Eqs (42), (43)
and
(44) are written in terms of two independent solutions
of Eqs. (19), (20) and (21) respectively. These
solutions are
\begin{equation}
Z(r)=\hbar \arctan \left\{ {\frac{\mu
_{r}R_{1}(r)+R_{2}(r)}{R_{1}(r)+\nu
_{r}R_{2}(r)}}\right\} \; ,
\end{equation}
\begin{equation}
L(\vartheta )=\hbar \arctan \left\{ {\mu _{\vartheta }
T_{1}(\vartheta)+T_{2}(\vartheta ) \over
T_{1}(\vartheta )+\nu _{\vartheta }T_{2}(\vartheta
)}\right\} \; ,
\end{equation}
\begin{equation}
M(\varphi )=\hbar \arctan \left\{ {\frac{\mu _{\varphi
}\sin (m_{\ell
}\varphi )+\cos (m_{\ell }\varphi )}{\sin (m_{\ell
}\varphi )+\nu _{\varphi
}\cos (m_{\ell }\varphi )}}\right\} \;,
\end{equation}
in which we have used Bouda's notation \cite{Bouda},
and where
$\mu _{r}$, $\nu _{r}$, $\mu _{\vartheta }$, $\nu
_{\vartheta }$, $\mu
_{\varphi }$ and $\nu _{\varphi }$ are real constants.
$R_{1}$
and $R_{2}$ are two real independent solutions of Eq.
(19).
$T_{1}$ and $T_{2}$ are two real independent solutions
of
Eq. (20). $\sin (m_{\ell }\varphi )$ and $\cos
(m_{\ell }\varphi )$ are two independent solutions of
Eq. (21).

%
\vskip\baselineskip
\noindent
{\bf 4\ \ The Cylindrical Symmetry Potential}
\vskip\baselineskip
%

Now, let us consider a potential with cylindrical
symmetry whose
expression is
\begin{equation}
V(\vec{r})=V(\rho)\; .
\end{equation}
The stationary Schr\"odinger equation written with the
cylindrical
coordinates $(\rho,\phi,z)$ \cite{Tannoudji} is
\begin{equation}
-{\frac{\hbar^2 }{2m}}\left({\frac{\partial^2 \psi
}{\partial \rho^2}}+ {%
\frac{1 }{\rho}}{\frac{\partial \psi }{\partial
\rho}}+ {\frac{1 }{\rho^2}}{%
\frac{\partial^2 \psi }{\partial \phi^2}}+
{\frac{\partial^2 \psi }{\partial
z^2}} \right)+V(\rho)=E\ \psi\; .
\end{equation}
In this case, we can write the wave function
$\psi(\rho,\phi,z)$ as
\begin{equation}
\psi(\rho,\phi,z)=G(\rho)\ N(\phi)\ U(z)\; .
\end{equation}
$G(\rho)$ is the radial wave function. $N(\phi)$ is
the angular wave
function while $U(z)$ is the axial wave function. By
substituting
relation (53) in Eq. (52), one obtain
\begin{equation}
-{\frac{\hbar^2 }{2m}}\left({\frac{d^2 G }{d\rho^2}}+
{\frac{1 }{\rho}}{%
\frac{dG }{d\rho}}\right)+ \left[ {\frac{m_{\phi}^2\
\hbar^2 }{2m \rho^2}}%
-E+V(\rho)- {\frac{\beta \ \hbar^2 }{2m
\rho^2}}\right]\ G(\rho)=0\; .
\end{equation}
\begin{equation}
{\frac{d^2 N }{d\phi^2}}+m_{\phi}^2 \ N(\phi)=0
\end{equation}
\begin{equation}
{\frac{d^2 U }{dz^2}}-\beta \ U(z)=0
\end{equation}
$\beta$ is a real constant and $m_{\phi}$ is an
integer one.
Eqs. (55) and (56) have the same form as the one of
Schr\"odinger
equation with vanishing potential, while Eq. (54) can
be written in the
Schr\"odinger equation form by setting
\begin{equation}
G(\rho)={\frac{H(\rho) }{\sqrt{\rho}}}\; .
\end{equation}
Taking twice the derivative of Eq. (57) with respect
to $\rho$, we get
to
\begin{equation}
{\frac{dG }{d\rho}}={\frac{dH/d\rho }{\sqrt{\rho}}}-
{\frac{H(\rho) }{%
2\rho^{\frac{3 }{2}}}}\; .
\end{equation}
\begin{equation}
{\frac{d^2 G }{d\rho^2}}={\frac{d^2 H/d\rho^2
}{\sqrt{\rho}}}- {\frac{%
dH/d\rho }{\rho^{\frac{3 }{2}}}}+ {\frac{3
}{4}}{\frac{H(\rho) }{\rho^{%
\frac{5 }{2}}}}\; .
\end{equation}
If we inject these two last relations in Eq. (54), we
deduce
\begin{equation}
-{\frac{\hbar^2 }{2m}}\ {\frac{d^2 H }{d\rho^2}}+
\left[ {\frac{%
(m_{\phi}^2\-1/4) \hbar^2 }{2m \rho^2}}- {\frac{\beta
\ \hbar^2 }{2m}}%
+V(\rho)\right]\ H(\rho)=E\ H(\rho)\; .
\end{equation}
Now, just as spherical symmetry and Cartesian symmetry
cases,
the functions $H(\rho)$, $N(\phi)$ and $U(z)$ take the
form (8)
\begin{equation}
H(\rho)=h(\rho)\left[\alpha_\rho\ e^{{\frac{i
}{\hbar}}Z_\rho(\rho)}+
\beta_\rho\ e^{-{\frac{i
}{\hbar}}Z_\rho(\rho)}\right]\; ,
\end{equation}
\begin{equation}
N(\phi)=n(\phi)\left[\alpha_\phi\ e^{{\frac{i
}{\hbar}}M_\phi(\phi)}+
\beta_\phi\ e^{-{\frac{i
}{\hbar}}M_\phi(\phi)}\right]\; ,
\end{equation}
\begin{equation}
U(z)=u(z)\left[\alpha_z\ e^{{\frac{i }{\hbar}}K(z)}+
\beta_z\ e^{-{\frac{i
}{\hbar}}K(z)}\right]\; ,
\end{equation}
where $\alpha_\rho$, $\beta_\rho$, $\alpha_\phi$,
$\beta_\phi$, $\alpha_z$ and $\beta_z$ are complex
constants. $h(\rho)$, $n(\phi)$ and $u(z)$ are real
functions of one variable. Injecting Eqs. (61), (62)
and
(63) in Eqs. (60), (55) and (56), we get, with same
procedure as in one dimension, to the following
equations
\begin{equation}
{\frac{1 }{2m}}\left( {\frac{dZ_\rho
}{d\rho}}\right)^2-{\frac{\hbar^2 }{4m}}%
\; \left \{Z_\rho, \rho \right \}+
V(\rho)+{\frac{(m_{\phi}^2\ -1/4) \hbar^2
}{2m \rho^2}}= E+{\frac{\beta \ \hbar^2 }{2m}}\; ,
\end{equation}
\begin{equation}
\left({\frac{dM_\phi }{d\phi}}\right)^2-{\frac{\hbar^2
}{4m}}\; \left
\{M_\phi, \phi \right \}=m_{\phi}^2\ \hbar^2\; ,
\end{equation}
\begin{equation}
\left( {\frac{dK }{dz}}\right)^2-{\frac{\hbar^2
}{4m}}\; \left\{ K,
z\right\}+\beta \ \hbar^2=0 \; ,
\end{equation}
Eqs. (64), (65) and (66) are the components of QSHJE
in 3-D for a potential with cylindrical symmetry. They
contain the Schwarzian derivatives of respectively
$Z_\rho$, $M_\phi$ and $K(z)$. By deducing the value
of $m_{\phi}$ from Eq. (65) and $\beta$ from (66) and
replacing in Eq. (64), we obtain
\begin{eqnarray}
{\frac{1 }{2m}}\left[\left({\frac{dZ_{\rho}
}{d\rho}}\right)^2+ {\frac{1 }{%
\rho^2}}\left({\frac{dM_\phi }{d\phi}}\right)^2+
\left({\frac{dK }{dz}}%
\right)^2 \right]- {\frac{\hbar^2 }{4m}}\; \left[\left
\{Z_\rho, \rho \right
\}+\right. \hskip5mm &&  \nonumber \\
\left. {\frac{1 }{\rho^2}}\left \{M_\phi, \phi \right
\}+ \left \{K,
z\right \}\right]+V(\rho)-{\frac{\hbar^2 }{8m\;
\rho^2}}=E\; .
\end{eqnarray}
Let us define the reduced action in 3-D for a
cylindrical
symmetry potential as
$$
S_0(\rho,\phi,z)=Z_\rho(\rho)+M_\phi(\phi)+K(z)\; .
$$
Injecting this last relation in Eq. (67), we get
\begin{equation}
{\frac{1
}{2m}}\left(\vec{\nabla}_{\rho,\phi,z}S_0\right)^2-
{\frac{\hbar^2
}{4m}}\; \left[\left \{S_0, \rho \right \}+ {\frac{1
}{\rho^2}}\left \{S_0,
\phi \right \}+ \left \{S_0, z\right
\}\right]+V(\rho)-{\frac{\hbar^2 }{8m\;
\rho^2}}=E\; .
\end{equation}
Eq. (68) represents the QSHJE in 3-D for a cylindrical
symmetry potential. At the classical limit $(\hbar \to
0)$,
Eq. (68) goes  to
\begin{equation}
{\frac{1
}{2m}}\left(\vec{\nabla}_{\rho,\phi,z}S_0\right)^2+
V(\rho)=E\; ,
\end{equation}
which is the CSHJE. As in the spherical symmetry case,
taking the classical limit in the three components of
the
QSHJE does not lead to the CSHJE because
the conjugate momenta $dM_\phi/d\phi$ and $dK/dz$
vanish.
The classical limit must be taken in Eq. (68).

Note that for this case, after having separate
variables in Eq. (68),
three constants of motion appear, $m_\phi$, $\beta$
and energy $E$. The constant $m_\phi$ will be linked,
in the following section, with the angular momentum
of the particle.

  %
\vskip\baselineskip
\noindent
{\bf 5- Trajectory Representation and the Spin}
\vskip\baselineskip
%

Now, let us make some comments on the results obtained
above.
First, remark that, for each of the three cases
studied in this paper,
quantum terms appear in the 3-D QSHJE. The Schwarzian
derivatives
appear for the three cases (Eqs. (15), (46) and (68)
), which means
that they take the fundamental role in the motion of
the particle.
These derivatives can have, for the dynamical motion,
the same role as the Schwarzian derivative  in one
dimension
(see Ref. \cite{BD1}).  For the spherical and
cylindrical symmetry
cases, we observe in addition of Schwarzian derivatives, 
more
quantum terms in the 3-D QSHJE.

\noindent
For the spherical symmetry potential, the QSHJE
(Eq. (46)) contains  two quantum terms
$$
ter_1=-{\hbar^2 \over 8mr^2}=-{(\pm 1/2)^2 \hbar^2
\over 2mr^2}
$$
 and
$$
ter_2 =-{\hbar^2 \over 8mr^2 \sin^2 \vartheta}
=-{(\pm 1/2)^2 \hbar^2 \over 8mr^2} \; .
$$
These two terms are purely quantum's, since at the
classical limit they vanish. $ter_1$ has, as we can
see  in Eq. (42), the form of $-\lambda \hbar^2 
/2mr^2$ in which
$\lambda=\ell (\ell +1)$ represents the eighen value
of $L^2$
($L$ being angular momentum operator). For these
reasons, it is obvious to link the quantity
$1/4=(\pm 1/2)^2$ appearing in $ter_1$ with a residual
angular momentum which may be connected, as a first
approach, with the spin momentum of the particle.
This can be done since the semi-integer value $1/2$
appears
only for spin momentum.  The term $ter_2$ has the form
of
$m_{\ell}^2 /2mr^2 \sin^2 \vartheta$    in which
$m_\ell$ is the
eighen value of $L_z$. This suggest that  the
quantities
$(\pm 1/2)^2$ are related to the two projections of
the residual
angular momentum or spin. Otherwise, the quantum terms
$(-\hbar^2 /8mr^2)$ and $(-\hbar^2 /8mr^2 \sin^2
\vartheta)$
in Eq. (46) can be seen, in a first approach, as the
manifestation of the spin momentum. And, the vanishing
of these terms at the classical limit $(\hbar \to 0)$
is in
agreement with the fact that the Spin is a quantum
characteristic of particles.

\noindent
The cylindrical symmetry case shows the same
property of residual angular momentum (Spin) since a
quantum
term $-\hbar^2 / 8m\rho^2$ appears in the 3-D QSHJE
(Eq. (68)).

{\bf Conclusions}
In the present article, we have introduced a
generalization of
the one dimensional QSHJE -introduced and investigated
by
Floyd \cite{Floyd1,Floyd2,Floyd3,Floyd5}- into three
dimensions
for three cases, Cartesian, spherical and cylindrical
symmetry
potentials. The aim of such as generalization is to
make
possible the study of three dimensional motions.
Hence,
Eqs. (15), (46) and (68) can be considered as the
first step
of a dynamical approach in 3-D. For the Cartesian
symmetry case,
such approach is obvious, since we can separate the
three
directions of motion, and define for each direction
the
corresponding Lagrangian in the same manner as it is
done in Ref. \cite{BD1}. For the other cases, it is
not obvious to
make a dynamical approach of the particle's motion.
But,
the presence of the Schwarzian derivatives in the 3-D
QSHJE
and the analogy between the form of the 3-D reduced 
actions
(Eqs. (46),  (47), (48)) and the one dimensional
reduced action \cite{Floyd2,Bouda} indicates that
such an
approach is possible.

Another point which is important is the fact that to
reduce Eqs.
(46) and (68) to the 3-D CSHJE, one must take the
limit in these
equations, not in the separated equations (Eqs. (42),
(43), (44),
(62), (63) and (64) respectively).

\vskip\baselineskip

\noindent
{\bf ACKNOWLEDGEMNTS}

I would like to think Dr. F. Djama for encouragement and help.

\newpage

%
\vskip\baselineskip
\noindent
{\bf REFERENCES}
\vskip\baselineskip
%

\end{document}